\documentclass[a4paper]{JHEP3}
\usepackage[T1]{fontenc}
\usepackage{psfrag,epsfig,subfigure}
\newcommand{\de}{\mathrm{d}}				
\newcommand{\ee}{\mathrm{e}}                

\title{Bouncing universe and non-BPS branes}

\preprint{LMU-ASC 26/09}

\author{Philipp
H\"offer~v.~Loewenfeld$^{1,}$\thanks{von.Loewenfeld@physik.lmu.de},
Jin~U Kang$^{1,2,}$\thanks{Jin.U.Kang@physik.lmu.de},
Nicolas Moeller$^{1,}$\thanks{Nicolas.Moeller@physik.lmu.de},
Ivo Sachs$^{1,}$\thanks{Ivo.Sachs@physik.lmu.de}\\ 
$^1$Arnold Sommerfeld Center for Theoretical Physics (ASC),
Ludwig-Maximilians-Universit\"at M\"unchen, Theresienstr.\ 37, 80333
M\"unchen, Germany\\
$^2$Department of Physics, Kim Il Sung University, Pyongyang, DPR.\
Korea}

\abstract{ We describe string frame bouncing universe scenarios
  involving the creation and annihilation of a non-BPS D9-brane in
  type IIA superstring theory. We find several classes of solutions,
  in which the bounce is driven by the tachyon dynamics of the non-BPS
  brane. The metric and the dilaton are consistently described in
  terms of the lowest order effective action.  The bounce solutions
  interpolate between contracting and expanding pre-big bang (or
  post-big bang) solutions. The singular behavior of our bounce
  solutions is the same as that of the pre-big bang (or post-big bang)
  solution. Upon adding a simple dilaton potential the asymptotic
  curvature singularity is removed but the dilaton still grows without
  bound. Such a potential may result from $\alpha'$ corrections in the
  open string sector.  }

\keywords{string theory and cosmology, cosmic singularity}

\psfrag{F}{\(F\)}
\psfrag{W}{\(w\)}
\psfrag{X}{\(X\)}
\psfrag{Phi}{\(\Phi\)}
\psfrag{t}{\(t\)}
\psfrag{T}{\(T\)}
\psfrag{w}{\(w\)}
\psfrag{HE}{\(H_{\rm E}\)}
\psfrag{H}{\(H\)}

\begin{document}

\section{Introduction}\label{intro}
The resolution of the big-bang singularity is an important open
problem in standard cosmology (see e.\,g.\ \cite{Novello:2008ra,Brandenberger:2008nx} for a
review and references therein). At the same time it is a natural playground for string theory since quantum gravity corrections are expected to be relevant in this regime. In the ekpyrotic scenario \cite{Khoury:2001wf,Khoury:2001bz} and a 
refined version, the cyclic universe \cite{Steinhardt:2001st}, the hot 
big bang is the result of the collision of two branes. Explicit cyclic models have been suggested as effective four-dimensional models inspired from heterotic 
M-theory. The bounce in the ekpyrotic scenario occurs at a 
real curvature singularity and thus does not resolve the curvature singularity. 
The new ekpyrotic scenario \cite{Buchbinder:2007ad}, realizes an
explicit bounce dynamics by addition of a ghost condensate but suffers
from a vacuum instability problem (see \cite{Kallosh:2007ad}).
More generally, due to the necessary violation of the
null energy condition (NEC) during a bounce, phenomenological models
producing a bounce often suffer from the problem of introducing matter
with negative energy density, i.\,e.\ ghosts. 

There are other  ways to address the big-bang singularity problem in string 
theory\footnote{Alternative suggestions 
avoiding the problem of introducing matter with negative energy density include loop quantum cosmology
\cite{Ashtekar:2008ay} and matrix models
(e.\,g.\ \cite{Craps:2005wd,Klammer:2009ku}).}. The pre-big bang scenario (see \cite{Gasperini:2007vw} for a 
review) is a consequence of the fact that the tree-level equations of 
motion of string theory are not only symmetric under time reflection 
\(t\mapsto -t\) but also symmetric 
under the scale-factor duality transformation \(a \mapsto \frac1a\) with 
an appropriate transformation of the dilaton. The `post-big bang' 
solution of standard cosmology with decelerated expansion defined for 
positive times is by these dualities connected to an inflationary 
`pre-big bang' solution for negative times. In this way the cosmic 
evolution is extended to times prior to the big bang in a self-dual way 
but the solution is still singular. One can obtain regular self-dual 
solutions by tuning a suitable potential for the dilaton. But albeit a 
potential of this form might be the result of higher-loop quantum 
corrections, its form has not been derived from string theory.


In this paper we consider a novel scenario in string theory where a
bounce occurs in the string frame due to the creation of an unstable
(non-BPS) brane as the universe bounces through a string size regime
before expanding as the brane decays. Our solutions effectively
interpolate either between a contracting and an expanding pre-big bang
solution or between a contracting and an expanding post-big bang
solution. The future (past) singularity of the pre (post)-big bang
solution is not resolved in this scenario.  The nice feature in our
model is that the curvature as well as the dilaton and its
derivative remain small (in string units) through the bounce so that
referring to perturbative string theory and the simplest low energy
effective action for these fields is justified. In addition no fine
tuning is required.  On the other hand, we do not address issues like
dilaton and moduli stabilization which are of course important
problems to embed this model into late time standard cosmology but are
not relevant during the string scale regime where the bounce occurs.
We should stress that our string frame bounce solutions describe
monotonously contracting or expanding geometries in the Einstein
frame. A bounce in the Einstein frame may occur upon stabilizing the dilaton
asymptotically. However, this entails violating the NEC or,
alternatively, allowing the gravitational coupling to change sign in
the string frame.  We will discuss a model where the latter effect
occurs.

Our work is organized as follows. This paper consists of seven sections,
of which this introduction is the first. The model we employ is
described in Section~\ref{SectionModel}. In this section we present
the effective actions and the equations of motions for the metric,
dilaton and tachyon from a non-BPS brane, which makes the bounce
possible. In Section~\ref{SectionFrame} we show that our model has no
ghost by considering the null energy condition in the Einstein frame,
and some features of string frame bounce scenarios are studied in
relation to the Einstein frame. In Section~\ref{SectionAsymptotic},
the asymptotic behavior of the solutions is analyzed and its
qualitative similarity to pre-big bang scenario is clarified. The
numerical determination of the global bounce solution is presented in
Section~\ref{SectionNumerical}. In Section~\ref{SectionNonsingular} we
present a simple model that resolves the asymptotic curvature
singularity in the string frame. Finally we conclude in the last
section.

\section{The model}\label{SectionModel}
We consider a non-BPS space-filling D9-brane in type IIA superstring
theory. The details of the
compactification will not play a role here. Concretely we consider the
lowest order effective action for the metric and dilaton in the string
frame as well as an effective action for the open tachyonic mode of
the non-BPS D-brane. For now, the only assumption being made for the
tachyon action is that only the first derivatives of the tachyon
appear in it. The ansatz for the gravitational action is justified
provided the dilaton and metric are slowly varying in string units.
We write
\begin{eqnarray}
  S &=& \frac{1}{2\kappa_{10}^{2}}\int\de^{10}x \, \sqrt{-g} \, \ee^{-2\Phi}
\left(R+4 \, \partial_{\mu}\Phi\partial^{\mu}\Phi\right) + S_{\rm T}
\label{eq:original action} \\
\hbox{with} \quad S_{\rm T} &=& \int\de^{10}x \, \sqrt{-g} \, \ee^{-\Phi}
L(T, \partial_{\mu} T \partial^{\mu} T),
\label{eq:original action-2}
\end{eqnarray}
where $\Phi$ is the dilaton, $T$ is the tachyon, and $\kappa_{10}^2 =
8 \pi G_{10}$ with $G_{10}$ the ten-dimensional Newton constant. We
use the signature $(-,+,\ldots,+)$ for the metric. With these
conventions, the matter energy-momentum tensor is given by
\begin{equation}
T^{\mu}_{\nu} = \frac{2}{\sqrt{-g}} \, \frac{\delta S_{\rm T}(T,\partial_{\rho} T 
\partial^\rho T, \Phi)}{\delta g_{\mu}^{\nu}}.
\end{equation}
For a Lagrangian minimally coupled to gravity, the metric appears only
in $(\partial T)^2 = \partial_{\rho} T \partial^{\rho} T$, and we can
write the energy-momentum tensor as
\begin{equation}
T^{\mu}_{\nu} = 2 \, \ee^{-\Phi} \, \frac{\partial L(T, (\partial T)^2)}
{\partial \left( (\partial T)^2 \right)} \, \partial^{\mu} T\partial_{\nu} T - 
\delta^{\mu}_{\nu} \, \ee^{-\Phi} L \label{tmunu}
\end{equation}
In a homogeneous isotropic universe that we will consider, all fields
are assumed to depend only on time; the energy density $\epsilon$ and
pressure $p$ are given by
\begin{equation}
\epsilon = T_0^0 \quad , \qquad p = -T_i^i \ \hbox{(no sum)} \ = \ee^{-\Phi} L.
\label{energypressure}
\end{equation}

We are now ready to make an ansatz for the metric. We take a
four-dimensional spatially flat FRW spacetime times a six-torus characterized by a
single modulus $\sigma$. Namely
\begin{equation}
\de s^2 = -\de t^2 + a(t)^2 \, \delta_{ij} \de x^i \de x^j + \ee^{2\sigma(t)} \, \delta_{IJ} \de x^I \de x^J,
\label{ansatz}
\end{equation}
where the lower-case Latin indices run over the three uncompactified
space coordinates, while the upper-case Latin indices label the six
compactified dimensions. We further simplify the problem by
restricting to the case where $a(t) = \ee^{\sigma(t)}$.
Although equality of the two scale factors is phenomenologically not
satisfying at late times it may be assumed near the cosmological
bounce which is the prime focus of this paper. In particular, we will
not address the important problem of moduli- and dilaton stabilization
required to connect to the standard cosmology at late times. With the
ansatz (\ref{ansatz}) the Einstein equations are now effectively
isotropic.  And in particular the relations (\ref{energypressure}) can
be used. From now on the indices $i, j$ include $I, J$.

The equations of motion for $g_0^0$, $g_i^i$ and $\Phi$ are then given
by the following first three equations
\begin{eqnarray}
72H^{2} - 36H\dot{\Phi} +4\dot{\Phi}^{2} - 2 \kappa_{10}^2 \, \ee^{2\Phi} \, \epsilon&=&0,\label{eq:eom-T-FRW-9g10}\\
2\ddot{\Phi}-8\dot{H} +16H\dot{\Phi} -2\dot{\Phi}^{2} -36H^2 - \kappa_{10}^2 \, \ee^{2\Phi} \, p&=&0,
\label{eq:eom-T-FRW-9g11}\\
2\ddot{\Phi} +18H\dot{\Phi} -2\dot{\Phi}^{2}-9\dot H-45H^2 - 
\frac{\kappa_{10}^2}{2} \, \ee^{2\Phi} \, p &=&0,
\label{eq:eom-T-FRW-9g12}\\
\dot\epsilon+9H(\epsilon+p)-\dot\Phi \, p &=&0,
\label{eq:eom-T-FRW-9g}
\end{eqnarray}
and the last equation follows from the generalized conservation law 
$\nabla_{\mu} T^{\mu}_{\nu} = (\partial_{\nu}\Phi) \, \ee^{-\Phi} \, L$. 

In the case where $\epsilon = 0$ and $p=0$, these equations allow
exact solutions, namely $H=\pm\frac{1}{3t}$,
$\dot{\Phi}=\frac{-1\pm3}{2t}$.  These are the special cases of the
pre -big bang ($t<0$) and post-big bang ($t>0$) solutions, which
respect the time reflection symmetry ($t\mapsto-t$) and scale-factor duality
symmetry ($a\mapsto a^{-1}$) (see for example \cite{Gasperini:2007}).

Let us explore the possibility of the bounce. Subtracting
Eq.~(\ref{eq:eom-T-FRW-9g12}) from Eq.~(\ref{eq:eom-T-FRW-9g11}) we
find
\begin{equation}
\dot{H} +9H^2-2H\dot{\Phi} - \frac{\kappa_{10}^2}{2} \, \ee^{2\Phi} \, p = 0\,.
\label{eq:eom-T-FRW-9H}
\end{equation}
This is an important equation. It tells us that a necessary condition
to have a bounce,
\begin{equation}
H=0 \qquad \hbox{and} \qquad \dot H > 0\,,
\end{equation}
is that the tachyon pressure $p$ must be {\em positive}. This is a
tight constraint for a scalar field action; the Born-Infeld action for
instance, which is an often used ansatz as a higher derivative scalar
field action, gives a pressure that is always negative.

Furthermore, assuming that $H$ is negative during a contracting phase
with growing dilaton and a negative pressure, (\ref{eq:eom-T-FRW-9H})
implies $\dot H<0$, i.\,e.\ accelerated contraction. On the other hand,
for positive equation of state for the scalar field, $w>0$, the
conservation equation in (\ref{eq:eom-T-FRW-9g}) gives a growing
pressure $p$ in the contracting phase so that a ``turn around'' $\dot
H=0$ is compatible with (\ref{eq:eom-T-FRW-9H}).

Of course, a Born-Infeld action for the tachyonic sector of non-BPS
branes is not in any way suggested by string theory. On the other
hand, within the restriction to first derivative actions it is
possible to derive an approximate effective action from string theory
for the open string tachyon of an unstable brane. This action,
constructed in \cite{LS1} and further studied in \cite{LS2} is given
by
\begin{equation}
L = -\sqrt{2} \, \tau_9 \, \ee^{-\frac{T^2}{2 \alpha'}} \left( \ee^{-(\partial T)^2} + 
\sqrt{\pi (\partial T)^2} \, \mathop{\rm erf}\left(\sqrt{(\partial T)^2} \right) \right),
\label{erf-action}
\end{equation}
where $\tau_9$ is the tension of a BPS 9-brane, and therefore
$\sqrt{2} \, \tau_9$ is the tension of a non-BPS 9-brane
\cite{Sen:1999md}. Let us shortly summarize how this action was
constructed. First, setting $(\partial T)^2$ to zero, we see that the
potential is given by
\begin{equation}
V(T) = \sqrt{2} \, \tau_9 \, \ee^{-\frac{T^2}{2 \alpha'}}\,.
\end{equation}
This is the exact potential for the open string tachyon potential
found in boundary superstring field theory \cite{BSFT1, BSFT2, BSFT3}.
The locations of the minima of $V(T)$ are at $T=\pm\infty$. At these
values the energy is degenerate with the closed string vacuum which
means that the non-BPS brane is absent. The construction of the full
action (\ref{erf-action}) is based on the observation that the tachyon
kink $T(x)=\chi \, \sin(x / \sqrt{2 \alpha'})$, where $x$ is one of
the spatial world volume coordinates, is an exactly marginal
deformation of the underlying boundary conformal field theory
\cite{Callan:1994ub, Recknagel:1998ih} and thus should be a solution
of the equations of motion obtained from (\ref{erf-action}). It turns
out that this requirement determines uniquely the action once the
potential has been chosen.  Furthermore, it follows by analytic
continuation that Sen's rolling tachyon solution \cite{Sen:rolling}
\[T(t) = A \, \sinh(t / \sqrt{2 \alpha'}) + B \, \cosh(t / \sqrt{2
  \alpha'})\] is also a solution of the action (\ref{erf-action}) for
all values of $A$ and $B$. In this dynamical decay (or creation) of
the non-BPS brane the energy is conserved. The asymptotic state for
large positive (or negative) times has been argued to be given by
"tachyon matter" - essentially cold dust made from very massive closed
string states.

\DOUBLEFIGURE{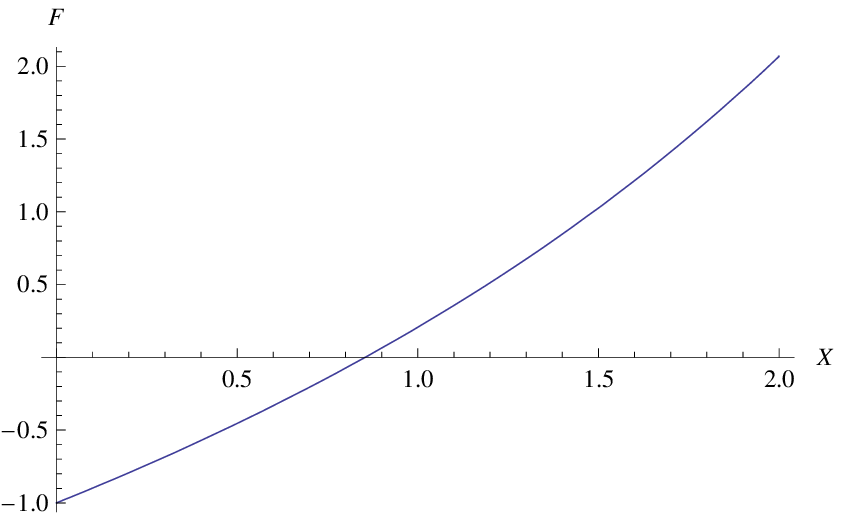,width=.475\textwidth}{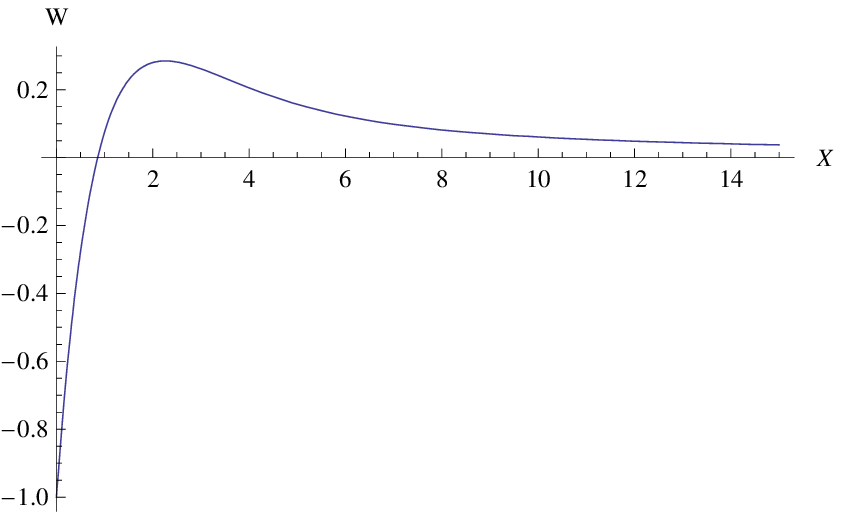,width=.475\textwidth}{$X$-dependence of the
  pressure $F(X)$.\label{Pressure}}{Equation of state \(w\) of the tachyon as a function of $X$.\label{W-efn-action}}
Let us now briefly explain how this action can allow a positive
pressure, or equivalently a positive Lagrangian \cite{LS2}. This
follows from the fact that it is real and continuous also for negative
 values of $(\partial T)^2$. Indeed if we write $-(\partial T)^2 = X$
(note that $X = \dot{T}^2$ in the homogeneous case), we can see that
\begin{equation}
\sqrt{- \pi X} \, \mathop{\rm erf}(\sqrt{-X}) = -2 \, \sqrt{X} \int_0^{\sqrt{X}} \ee^{s^2}\,\de s\,.
\end{equation}
This is negative and grows in absolute value faster than the first
term $\ee^X$ in the Lagrangian; so for positive enough $X$ (for
negative enough $(\partial T)^2$), the Lagrangian, and thus the
pressure, is always positive. This can be seen from
Fig.~\ref{Pressure}, where we show the $X$ - dependence of the
pressure, $F(X)\equiv-\left(\ee^{X}-2\sqrt{X}\int_0^{\sqrt{X}}\ee^{s^2}\,\de s\right)$. 
In terms of $F$ the Lagrangian can be expressed as $L =\sqrt{2}\tau_9 \ee^{-\frac{T^2}{2\alpha'}} F(X)$.  
From the figure it is clear that $\de F/\de X>0$ 
(this is also clear from $\frac{\de\,F}{\de X}=\frac{1}{\sqrt{X}}\int_0^{\sqrt{X}}\ee^{s^2}\,\de s >0$). 
From now on we work in the unit system with $\alpha'=1/2$. 
The energy density and pressure of the tachyon are 
\begin{equation}
\epsilon = \sqrt{2} \tau_9 \ee^{-\Phi}\ee^{-T^2+\dot{T}^2} \quad , \qquad p = \sqrt{2} \tau_9 \ee^{-\Phi}\ee^{-T^2}F(\dot{T}^2)\,.
\label{energypressure-1}
\end{equation}
Note that the equation of state $w=\epsilon/p$ depends only on
$\dot{T}^2$ as shown in Fig.~\ref{W-efn-action}. In particular, $w\to 0$ as $\dot{T}^2\to \infty$, while $w\to -1$ as $\dot{T}^2\to 0$. 

\section{Einstein frame and null energy condition}\label{SectionFrame}
In this section we show that our model has no ghost and satisfies 
NEC in the Einstein frame. This is important because there
might be a pathology due to an instability coming from NEC violation.  
On the other hand,
the consideration on NEC will help us draw more general conclusions
concerning the bounce scenarios.

The action (\ref{eq:original action})-(\ref{eq:original action-2}) 
in the string frame with $L$ given in (\ref{erf-action}) is expressed
in the Einstein frame by means of a conformal transformation
$g_{\mu\nu}=\tilde{g}_{\mu\nu}\, \ee^{\frac{\Phi}{2}}$ (where
$\tilde{g}_{\mu\nu}$ is the metric in the Einstein frame). This yields
\begin{eqnarray}
 S &=& \frac{1}{2\kappa_{10}^{2}}\int\de^{10}x \, \sqrt{-\tilde{g}} \, 
\left(\tilde{R}- \frac{(\tilde{\nabla}\Phi)^2}{2}\right) + S_{\rm T} 
\label{eq:original action-Einstein} \\
\hbox{with} \quad S_{\rm T} &=& \int\de^{10}x \, \sqrt{-\tilde{g}} \, \ee^\frac{3\Phi}{2}
L(T, \tilde{X} \ee^{-\frac{\Phi}{2}})\,,
\label{eq:original action-2-Einstein}
\end{eqnarray}
where $\tilde{R}$ and $\tilde{\nabla}$ are the scalar curvature and
the covariant derivative associated with $\tilde{g}$, and
$\tilde{X}=-(\tilde{\nabla} T)^2=X \ee^\frac{\Phi}{2}$.

The relevant quantity for verifying classical stability and
nonexistence of ghost is the sign of the slope of the kinetic term
with respect to the first derivative of the fields, which should be
positive. If it is negative, it signals the existence of a ghost
(quantum mechanical vacuum instability) and also that the squared speed
of sound is negative (classical instability) \cite{Kallosh:2007ad}.
Since the dilaton has the correct sign for the kinetic term in the
Einstein frame, the only possible source of ghost is the tachyon
Lagrangian.  Therefore, one should only check the sign of
$\frac{\de\,L}{\de {\tilde X}}$. We have that
$\frac{\de\,L}{\de \tilde{X}}=\frac{\de\,L}{\de X}\frac{\de\,X}{\de\tilde{X}} \sim
\ee^{\frac{\Phi}{2}} \frac{\de\,F}{\de X}=
\frac{\ee^{\frac{\Phi}{2}}}{\sqrt{X}}\int_0^{\sqrt{X}}\ee^{s^2}\,\de s>0$.
So there is no ghost and therefore no violation of NEC in our model.

Now we turn to the issue of the bounce. In the spatially flat FRW
spacetime we are considering, the Hubble parameter $H_{\rm E}$ in the
Einstein frame is related to that in the string frame by
$H_{\rm E}=H-\dot{\Phi}/4$. Since the null energy condition is not
violated in our model, the Hubble parameter in the Einstein frame
monotonously decreases, so that the bounce cannot arise in the
Einstein frame. If the dilaton were frozen, it would be impossible to
have a bounce in the string frame as well because the two frames would
be trivially related (in particular $H_{\rm E}=H$). This is why the
running  dilaton is crucial for the string frame bounce. Now we make some remarks on the bounce scenario in the
string frame. We assume that the bounce arises as an interpolation
between two out of four different phases (i.\,e.\ contracting or expanding pre/post-big bang phases) 
in the pre-big bang scenario. Four transitions are
then possible, namely from the contracting pre-big bang phase to
expanding pre/post-big bang, or from contracting post-big bang to
expanding post/pre-big bang. But the transition from pre-big bang to
post-big bang cannot happen in our model because the NEC in the
corresponding Einstein frame is not violated. To show this, we note
that in the pre-big bang scenario with $S_{\rm T}=0$ the solutions are
\begin{equation}
H=\frac{n}{t-t_0} \quad , \qquad \dot{\Phi}=\frac{9n-1}{2(t-t_0)} 
\quad \hbox{with} \quad n=\pm\frac{1}{3}\,.
\label{pre-big-bang}
\end{equation}
Here, $t<t_0$ corresponds to the pre-big bang phase, and $t>t_0$ to
the post-big bang phase. It then follows that $H_{\rm E}=\frac{1-n}{8(t-t_0)}$ is
negative for pre-big bang solutions and positive for post-big bang
solutions, meaning that pre/post-big bang solutions correspond to a
contracting/expanding universe in the Einstein frame, regardless of
whether the universe is contracting or expanding in the string frame.
This means that a transition from pre-big bang to post-big bang in the
string frame corresponds to a bounce in the Einstein frame, which is
impossible unless the NEC is violated. Therefore, under the assumption
mentioned above, the only possible string frame bounce scenarios 
in our model are the interpolations either between two
pre-big bang phases or between two post-big bang phases.

This argument can be extended to the case where the asymptotic
behavior of the solutions in the string frame is qualitatively, but
not exactly, in agreement with the pre-big bang scenario, This is the
case in our model, as we will see in the next section. In conclusion,
if one is given a model that does not violate the NEC and if one knows
the asymptotic boundary conditions of the solution in the string
frame, one can then predict the possible bouncing scenarios in the
string frame by looking at the corresponding Einstein frame. An
example will be given in the next section.

\section{Asymptotic analysis}\label{SectionAsymptotic}
In this section we will try to obtain approximate analytic solutions.
This will provide us with the asymptotic boundary conditions for the
numerical solutions in the next section. We emphasize here that by
``asymptotic'' we mean $t \rightarrow - \infty$, or $t$
approaching a pole $t_0$, at which $H$ diverges, as in the pre-big bang scenario. Similarly the present analysis applies to $t \rightarrow  \infty$, or $t$
approaching a pole $t_0$, at which $H$ diverges as in a post-big bang scenario 

To simplify our
analysis, we will assume that in either of these limits, the tachyon
behaves like dust, i.\,e.\ $p=w(t)\epsilon$ with $w(t)\to 0$. This is
equivalent to claiming that $|\dot{T}|\to\infty$ asymptotically
because $p\propto \epsilon/(\dot T)^2$ when $|\dot{T}|\to\infty$ (see
\cite{LS2} for details). To justify this assumption, we look at the
tachyon equation of motion following from (\ref{erf-action}). We have
\begin{equation}
\ddot T + \left( 9H - \dot \Phi \right) D(\dot T) - T = 0\,,
\label{TachyonEoM}
\end{equation}
where the function $D(y) = \ee^{-y^2} \int_0^y \ee^{s^2}\de s$ is known
as the {\em Dawson integral}. This function is an odd function, and thus vanishes
at $y=0$. We will also use the fact that $D(y) = \frac{1}{2 y} +
{\cal O}(y^{-3})$ for $|y| \rightarrow \infty$. For $|t| \rightarrow
\infty$, we will see that $(9H - \dot \Phi)$ tends to zero. We can
thus ignore the second term in the equation of motion\footnote{Note that  we ignore
the possibility $T \rightarrow 0$ in the infinite past or
future; we are only interested in the cases where the D-brane is
absent in these limits.} 
(\ref{TachyonEoM}). Thus $\ddot T = T$ and
$T$, as well as $\dot T$, will grow exponentially; and the pressure
will thus vanish for $|t| \rightarrow \infty$. We emphasize that since
the pressure vanishes {\em exponentially fast}, we can simply remove
it from the asymptotic equations of motion because it will always be dominated by
the other terms that will vanish with a power law.

When $t$ approaches a pole $t_0$ the analysis is slightly different
because there the term $(9H - \dot \Phi)$ diverges (unless $\dot \Phi
\sim 9H$, but we will see that this does not happen on our numerical
solution). We will further assume that this term diverges at least as fast as
$\frac{1}{t-t_0}$. Assuming that $\dot T$ is finite at $t_0$ this then implies that either $T$ or
$\ddot T$ will diverge like $\frac{1}{t-t_0}$ or faster, in
contradiction with a finite $\dot T$. We cannot exclude the case where
$\dot T(t_0)$ vanishes in such a way that it precisely cancels the divergence in $(9H - \dot \Phi)$. In that case the second term of the equation of
motion could be regular at $t_0$, and thus $T$ and $\ddot T$ could be
regular there as well. We will nevertheless ignore this
possibility because $\dot T(t_0)$ corresponds to very particular initial
conditions. For generic initial conditions we will therefore have that
$|\dot T| \rightarrow \infty$ when $t \rightarrow t_0$. This then justifies our claim that we can ignore the pressure for the
asymptotic analysis. An immediate consequence of 
Eq.~(\ref{eq:eom-T-FRW-9g}) is then that $\epsilon \sim a^{-9}$.

We now will proceed by analyzing the system of equations
(\ref{eq:eom-T-FRW-9g10}-\ref{eq:eom-T-FRW-9g}) assuming that either
$\dot \Phi \propto H$, $|\dot \Phi| \ll |H|$ or $|\dot \Phi| \gg |H|$
asymptotically.

\begin{description}
\item[i)]
Let us first consider the possibility $\dot \Phi \propto H$. In that
case (\ref{eq:eom-T-FRW-9g11}) and (\ref{eq:eom-T-FRW-9g12}) imply
that either $|\dot H|\gg H^2$ or $ \dot H\propto H^2$. In the first
case we get $\dot \Phi \simeq 5 H$ and then (\ref{eq:eom-T-FRW-9g10})
implies that
\begin{equation}
 -8H^2=2\kappa_{10}^2\ee^{2\Phi}\epsilon
\label{eq:eom-T-FRW-9H32}
\end{equation}
i.e. negative energy. We thus exclude that possibility. In the second
case from (\ref{eq:eom-T-FRW-9g11}-\ref{eq:eom-T-FRW-9g12}) we obtain
the solutions of the pre-big bang scenario, i.\,e.\ Eq.~(\ref{pre-big-bang}). For consistency, we must verify that these
solutions satisfy the constraint (\ref{eq:eom-T-FRW-9g10}). This is
the case only when the energy density is subdominant compared to
the other terms in this equation. By using (\ref{pre-big-bang}), one
can see that $\ee^{2 \Phi} \epsilon$ goes like $\frac{1}{|t-t_0|}$ while 
the other terms behave like $\frac{1}{(t-t_0)^2}$, so the energy is
subdominant as $t\to t_0$. Thus the solutions can be approximated to
those of the pre-big bang scenario near the pole, $t_0$. At the same time we see that this possibility is excluded for $|t| \rightarrow \infty$.

\item[ii)]
  For $|\dot \Phi| \ll |H|$ Eqs.~(\ref{eq:eom-T-FRW-9g11}) and
  (\ref{eq:eom-T-FRW-9g12}) imply $27H^2+5\dot H=0$. On the other hand,
  setting $p=0$ in Eq.~(\ref{eq:eom-T-FRW-9H}) gives us $\dot{H} +9H^2
  = 0$, a clear contradiction. Thus, $|\dot \Phi| \ll |H|$ is excluded.

\item[iii)] We are thus left with the sole possibility $|\dot \Phi|
  \gg |H|$.  In that case (\ref{eq:eom-T-FRW-9g10}) implies
\begin{equation}
 2\dot \Phi^2=\kappa_{10}^2\ee^{2\Phi}\epsilon,
\label{eq:eom-T-FRW-9H52}
\end{equation}
and (\ref{eq:eom-T-FRW-9g11}) together with (\ref{eq:eom-T-FRW-9g12})
imply $-\ddot\Phi +\dot \Phi^2=0$, which gives $\Phi=-\log(|t-t_0|)$.
With $p = 0$, Eq.~(\ref{eq:eom-T-FRW-9H}) then implies $\dot{H}
-2H\dot\Phi=0$, which gives $H=\frac{h}{(t-t_0)^2}$, where $h$ is some constant. This is
consistent with Eq.~(\ref{eq:eom-T-FRW-9H52}) only for $|t| \to
\infty$ because $\ee^{2 \Phi} \epsilon \sim \frac{1}{t^2}$ as $|t| \to
\infty$.
\end{description}

To summarize, we find that the only
consistent asymptotic solution for $|t|\to \infty$ is given by
\begin{equation} \label{asympt}
\Phi\simeq -\log(|t|)\;,\qquad \qquad H\simeq\frac{h}{t^2}.
\end{equation}
and (\ref{pre-big-bang}) for $t \to t_0$.
Note that $H_{\rm E}\simeq -\frac{\dot{\Phi}}{4}$ for (\ref{asympt}).

Now we are in the position to predict the possible string frame bounce scenarios.
Following the same logic as in the last part of the previous section 
concerning the NEC in the Einstein frame, one can show that the only possible bounce scenario
is the transition either from pre-big bang-like solution\footnote{Note that what we refer to as 
pre/post-big bang-like solution is given in Eq.~(\ref{asympt}) with negative/positive time} to pre-big bang solution or 
from post-big bang solution to post-big bang-like solution. 
This is because the other transitions correspond to a bounce in Einstein frame, 
which is excluded in our model which satisfies the NEC.
To explain this in more detail, we consider the case where we start
with contracting pre-big bang-like phase.   
For large negative times our bounce solution is in agreement with the pre-big bang-like
solution with accelerated contraction of the universe and growing
dilaton.  Then the universe goes through a bounce and for $t\to t_0$
it approaches a pre-big bang solution with accelerated expansion and
growing dilaton. The Hubble parameter in the Einstein frame remains negative and keeps decreasing.
As will be seen in the next section, the numerical solutions are in good agreement with this picture. 

\section{Numerical results}\label{SectionNumerical}
In this section, we numerically solve
Eqs.~(\ref{eq:eom-T-FRW-9g10}-\ref{eq:eom-T-FRW-9g}) to obtain
global solutions. In what follows we set $2 \sqrt{2} \kappa_{10}^2
\tau_9=1$ (this can always be achieved by adding a suitable constant
to the dilaton), so that
\begin{equation}
2\kappa_{10}^2 \epsilon= \ee^{-\Phi} \ee^{-T^2+\dot{T}^2}
\end{equation}
\begin{equation}
2\kappa_{10}^2 \, p= -\ee^{-\Phi} \ee^{-T^2}\left(\ee^{\dot{T}^2}-2\sqrt{\dot{T}^2}\int_0^{\sqrt{\dot{T}^2}}\ee^{s^2} ds\right).
\end{equation}

\FIGURE{
\subfigure[Hubble parameter \(H\) (solid line)\label{Bounce-H}]{\epsfig{file=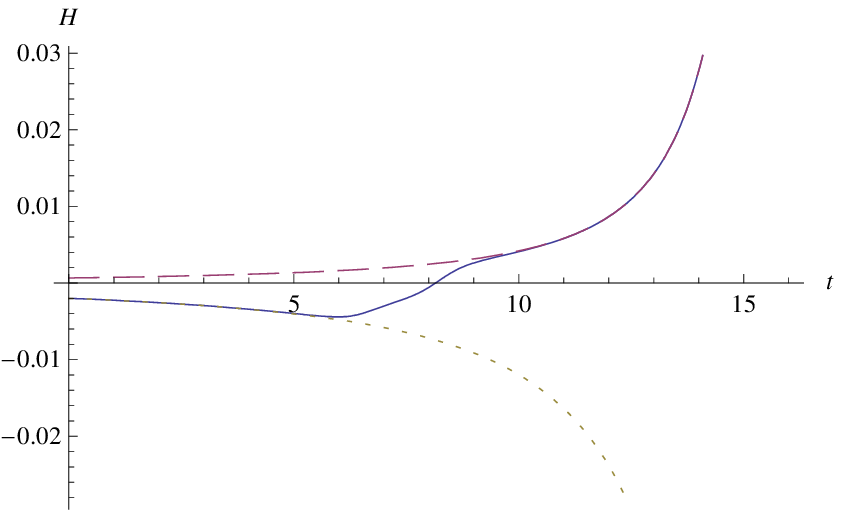,width=.475\textwidth}}
\subfigure[equation of state \(w\)\label{W1}]{\epsfig{file=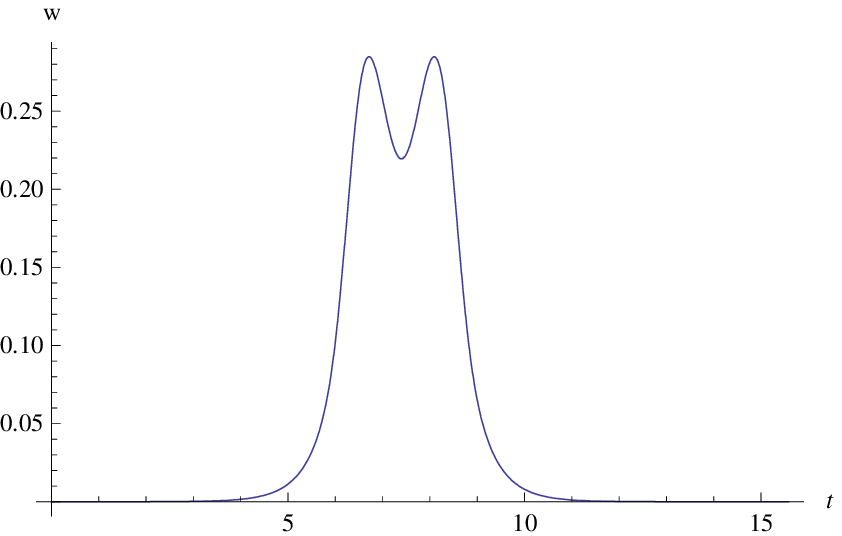,width=.475\textwidth}}
\subfigure[dilaton \(\Phi\)\label{Bounce-Phi}]{\epsfig{file=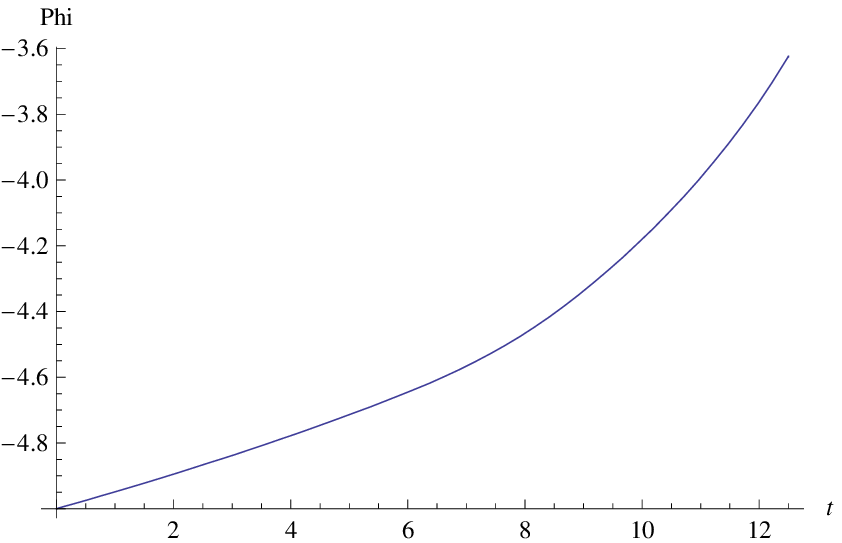,width=.475\textwidth}}
\subfigure[tachyon \(T\)\label{Bounce-T}]{\epsfig{file=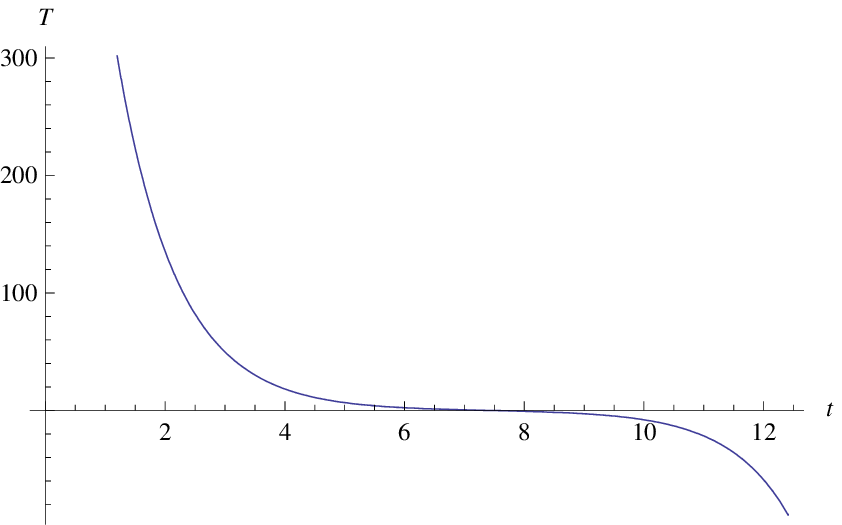,width=.475\textwidth}}
\caption{Bouncing numerical solution with the initial conditions $\Phi(0)=-5$, $\dot{\Phi}(0)=0.05$, $T(0)=1000$
and $H(0)= -0.002$.\label{Bounce-1}}}%
With this setup, we performed the numerical analysis and found a
family of bounce solutions. For example,
Figs.~\ref{Bounce-1} show a bounce solution with the
initial conditions, $\Phi(0)=-5$, $\dot{\Phi}(0)=0.05$, $T(0)=1000$
and $H(0)= -0.002$. The graph in Fig.~\ref{Bounce-H}a shows the evolution of the Hubble parameter (solid line). 
The bounce takes place near $t=8$. The bouncing solution can be seen as a transition from the
contracting pre-big bang-like phase (short dashed line) to the
expanding one (long dashed line). Both asymptotic solutions are
obtained by setting $p=0$ in the equations of motions since the
pressure is negligible in the far future and past (see Fig.~\ref{W1}b). The 'double bump' feature of the equation of state can be understood by noting that as the non-BPS brane builds up $|\dot T|$ decreases and thus $w$ increases from zero as explained in section 2. Then as $T$ reaches the top of the potential $|\dot T|$ becomes small and consequently $w$ decreases again. Indeed for $\dot T=0$ the equation of state is that of a cosmological constant. 

We found that a broad range of the initial conditions are allowed for
the bounce, so there is no fine-tuning problem. For instance,
$T(0)=\pm10^4$ and $T(0)=\pm100$ (keeping the same initial conditions
as above for the other variables) gives bounce solutions with
essentially the same behavior. Note that the large initial values for
$T$ do not represent a fine tuning. Rather it reflects the
condition that the non-BPS brane is absent at very early times. We see
that the asymptotic behavior of this family of solutions is similar to
the expanding pre-big bang case, in which the Hubble parameter blows
up. This agrees with the results of the previous section.

\FIGURE{
\subfigure[Hubble parameter \(H\) (solid line)\label{H-nonsingular-future}]{\epsfig{file=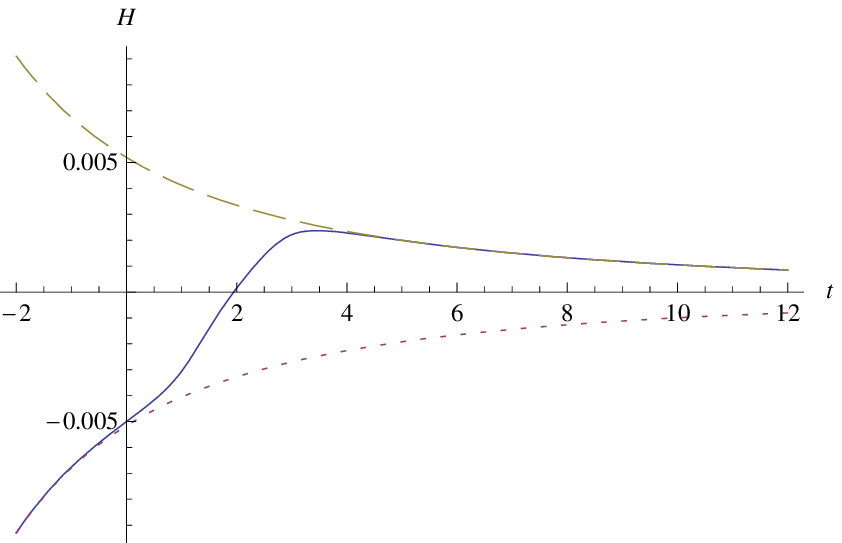,width=.475\textwidth}}
\subfigure[equation of state \(w\)\label{W2}]{\epsfig{file=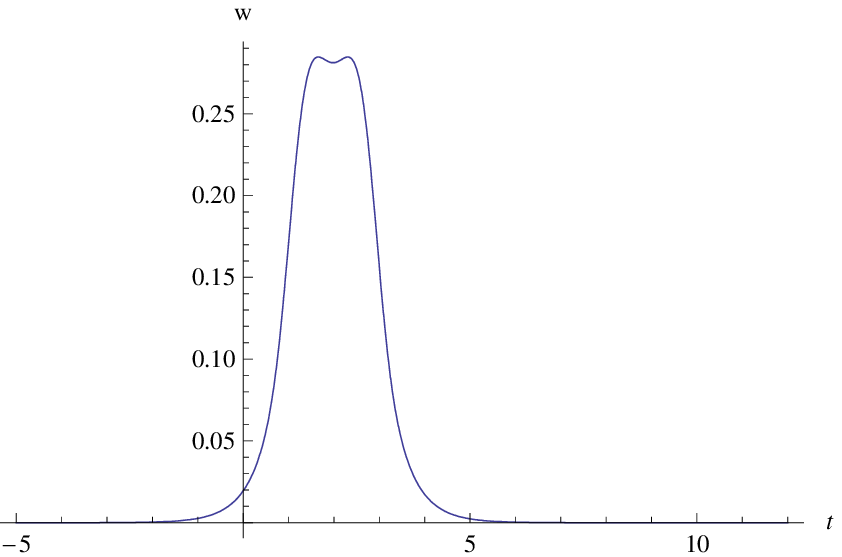,width=.475\textwidth}}
\subfigure[dilaton \(\Phi\)\label{Dilaton-nonsingular-future}]{\epsfig{file=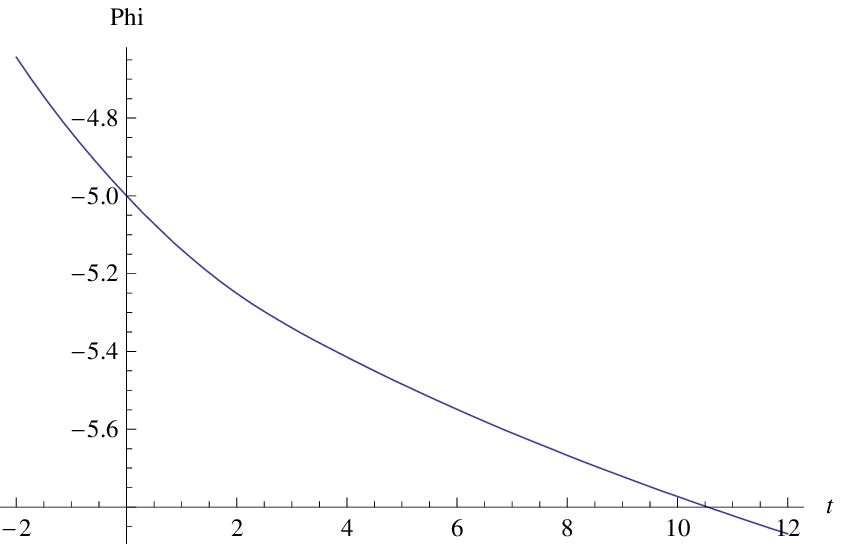,width=.475\textwidth}}
\subfigure[tachyon \(T\)\label{Tachyon-nonsingular-future}]{\epsfig{file=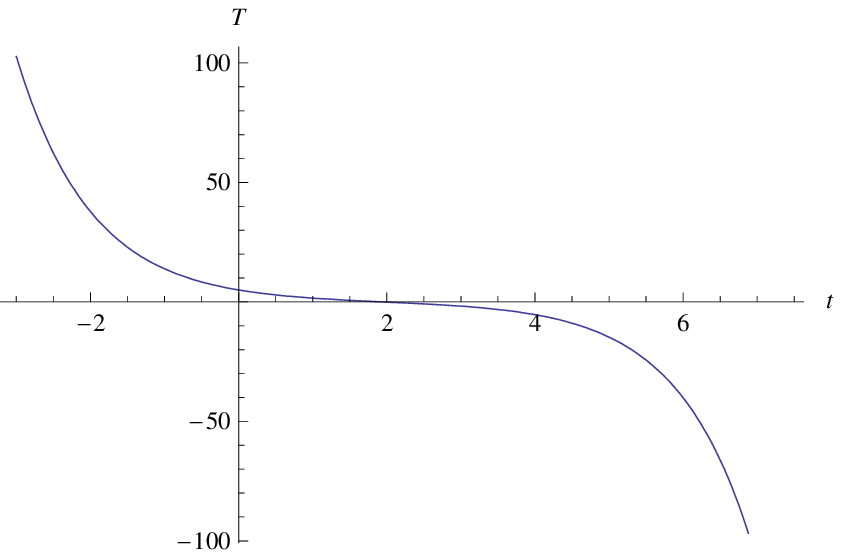,width=.475\textwidth}}
\caption{Bouncing, numerical solution with non-singular future using initial conditions $\Phi(0)=-5$,
  $\dot{\Phi}(0)=-0.15$, $T(0)=5$ and $H(0)= -0.005$.\label{nonsing-future}}}%
If one changes the sign of $\dot{\Phi}(0)$, a very different kind of
bounce is obtained. For instance,
Figs.~\ref{nonsing-future} show a bounce solution with the initial conditions, $\Phi(0)=-5$,
$\dot{\Phi}(0)=-0.15$, $T(0)=5$ and $H(0)= -0.005$. The graph in Fig.~\ref{H-nonsingular-future} shows the
  evolution of the Hubble parameter \(H\).  The bounce takes
  place near $t=2$, and the universe smoothly evolves to the standard
  cosmological regime, where the Hubble parameter and the dilaton decreases with time
(Fig.~\ref{Dilaton-nonsingular-future}). But going back in time
further, we found a singularity where the Hubble parameter blows up.  This
  solution can be seen as a transition from the contracting post-big
  bang phase (short dashed line) to the expanding post-big bang-like
  phase (long dashed line). Both asymptotic solutions are obtained by
  setting $p=0$ in the equations of motion since the pressure is
  negligible in the far future and past (see Fig.~\ref{W2}).

\FIGURE{\subfigure[Hubble parameter \(H\)\label{oscillatory-H}]{\epsfig{file=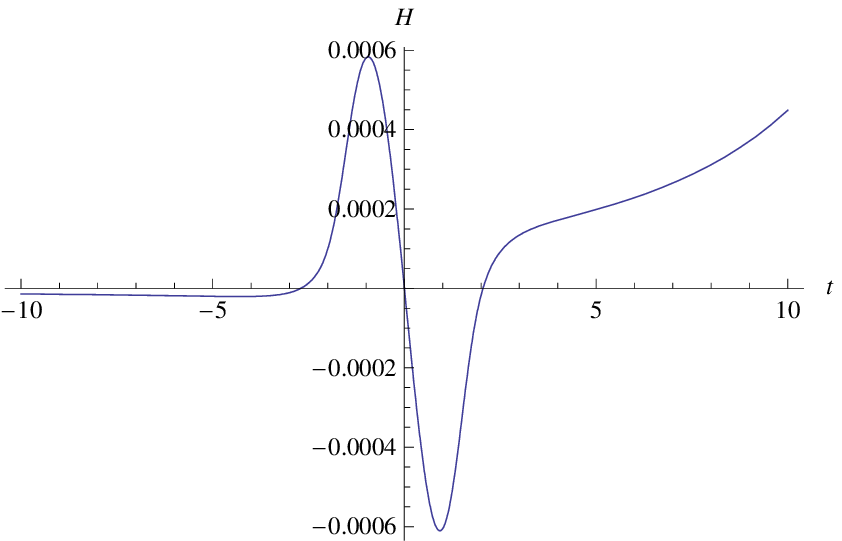,width=.475\textwidth}}
\subfigure[equation of state \(w\)\label{W3}]{\epsfig{file=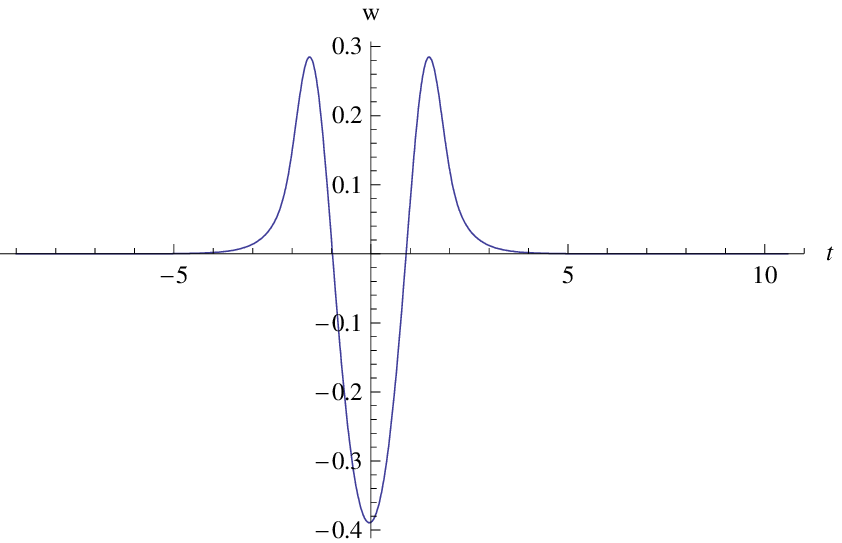,width=.475\textwidth}}
\caption{Oscillatory, numerical solution with initial conditions $\Phi(0)=-5$, $H(0)=0$,
$\dot{\Phi}(0)=0.05$ and $T(0)=0$.\label{oscillatory}}}%
\FIGURE{\subfigure[Hubble parameter \(H\)\label{oscillatory-H-nonsing-future}]{\epsfig{file=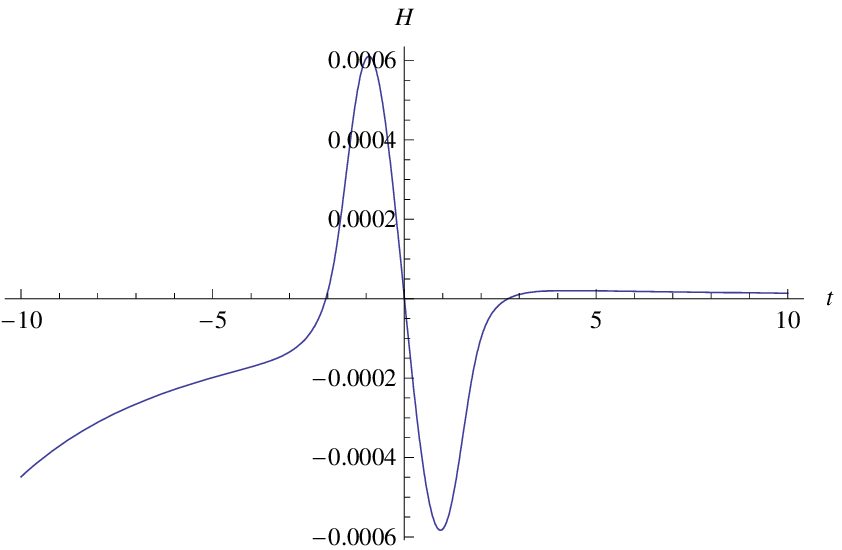,width=.475\textwidth}}
\subfigure[equation of state \(w\)\label{W4}]{\epsfig{file=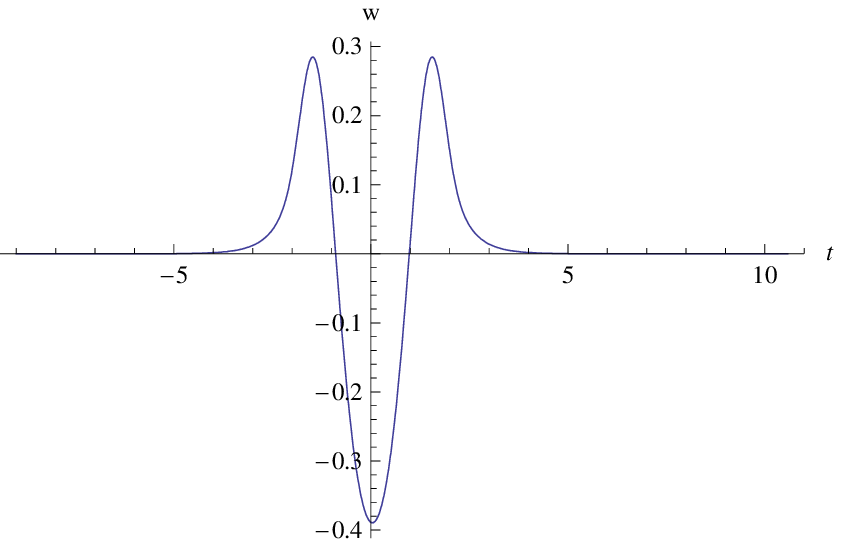,width=.475\textwidth}}
\caption{Oscillatory, numerical solution with initial conditions $\Phi(0)=-5$, $H(0)=0$, $\dot{\Phi}(0)=-0.05$ and $T(0)=0$.\label{oscillatory-nonsing-future}}}%
\FIGURE{\hspace{.2625\textwidth}\epsfig{file=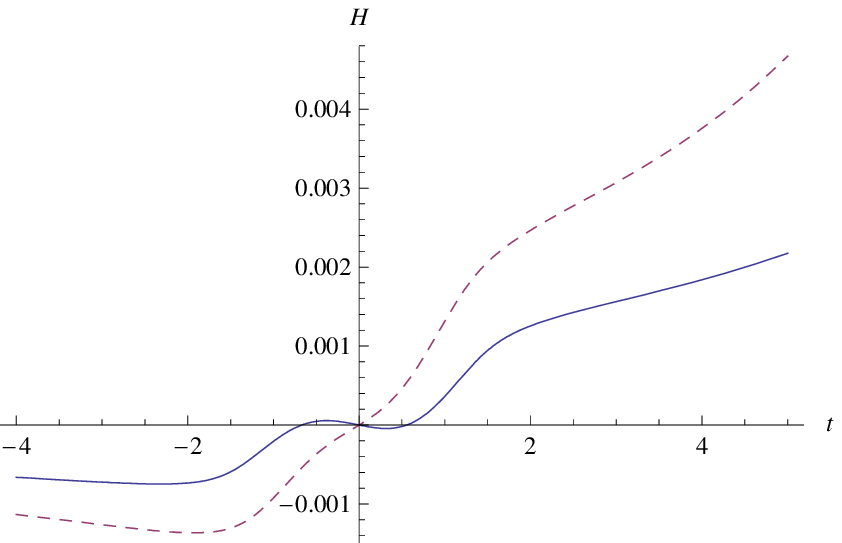,width=.475\textwidth}\hspace{.2625\textwidth}\caption{Hubble 
parameter \(H\) for $\dot{\Phi}(0)=0.06$ (solid
 line) and $\dot{\Phi}(0)=0.072$ (dashed line).\label{transitionToOscil}}}%
In addition, we found oscillatory solutions, in which a double bounce takes
place (see Figs.~\ref{oscillatory}). 
The solution in Figs.~\ref{oscillatory-nonsing-future} can be seen as a time
reflected one of the solution of Figs.~\ref{oscillatory}.  Which
solution is obtained depends on the sign of $\dot{\Phi}(0)$. In both
cases the asymptotic behaviors towards the curvature singularity are
analogous to the non-oscillatory cases mentioned above.
The evolutions of the equation of state are shown in Figs.~\ref{W3} and \ref{W4}.
Note that the negative equation of state implies small $|\dot{T}|$ (see Fig.~\ref{W-efn-action}), 
and this means that the oscillatory solutions can arise if the speed of the tachyon is small 
around the top of the tachyon potential (i.\,e.\ near $T=0$).
Alternatively 
this kind of solution can be obtained when we arrange $|\dot{\Phi}|$ 
to be small enough around the top of the tachyon potential (see
Fig.~\ref{transitionToOscil}). The variation of \(\dot{\Phi}(0)\) with the other 
initial conditions fixed (in this case at $\Phi(0)=-5$, $H(0)=0$, and $T(0)=0$)
shows that decreasing $\dot{\Phi}(0)$ gives rise to a transition from
single bounce to cyclic bounce.

In conclusion, we obtained bounce solutions, where the universe smoothly 
evolves from the contracting phase to the expansion. The bounce scenario 
that we here found can be classified into the following two cases. Note 
that these are exactly in agreement with our predictions on the bounce 
scenarios in the previous section: 
\begin{enumerate}
\item Transition from accelerating contraction (the contracting
  pre-big bang-like phase) to accelerating expansion (the pre-big bang
  inflation): In this case the dilaton grows up, and if the speed of
  the tachyon (or dilaton) is small enough near the maximum of the 
  tachyon potential, a double bounce can take place
  (Figs.~\ref{oscillatory}) before the universe evolves to the
  pre-big bang phase.
\item Transition from decelerating contraction (the contracting post-big bang phase)
 to decelerating expansion (post-big bang like
  phase): In this case the dilaton decays, and a double bounce can also
  take place (Figs.~\ref{oscillatory-nonsing-future}) under the same condition mentioned above.
\end{enumerate}

In all cases the tachyon rolls over the top of the potential in the
course of its evolution, and the bounce seems to happen when the
tachyon reaches around the top of the potential. The pressure is
important only around the bounce, and negligible (dust) asymptotically.

Our string frame bounce solutions correspond to monotonously
contracting (i.\,e.\ $H_{\rm E}=H-\frac{\dot{\Phi}}{4}<0$ ) or expanding
geometries ($H_{\rm E} > 0 $) in the Einstein frame (see Figs.~\ref{H-E-1}-\ref{H-E-2}), meaning that there is no bounce in
the Einstein frame for our solutions. This is the fact that our model
does not violate the NEC in the Einstein frame.
\DOUBLEFIGURE{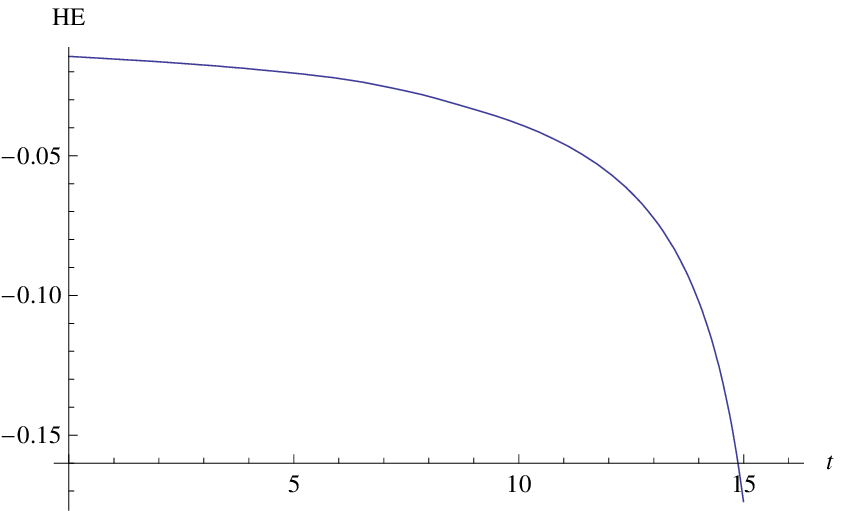,width=.475\textwidth}{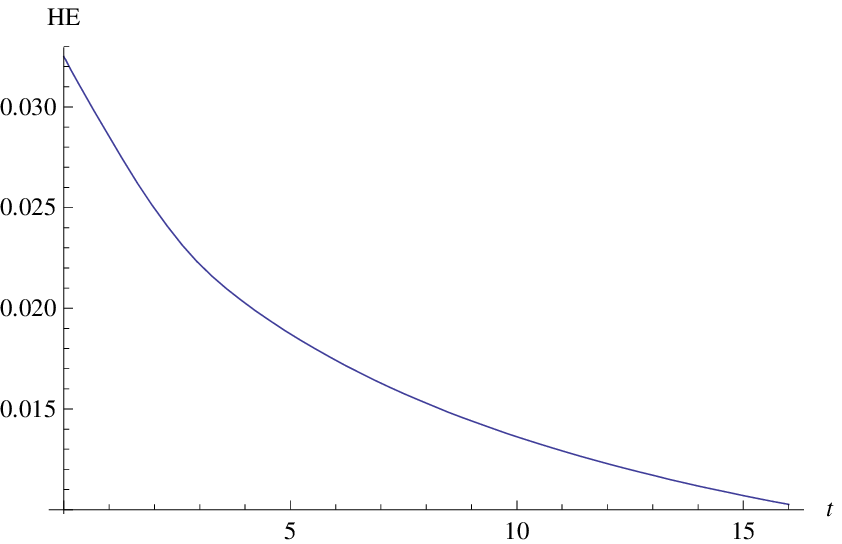,width=.475\textwidth}%
{Hubble parameter \(H_{\rm E}\) in the Einstein frame corresponding to Fig.~3~(a).\label{H-E-1}}
{Hubble parameter \(H_{\rm E}\) in the Einstein frame corresponding to Fig.~4~(a).\label{H-E-2}}

The numerical results presented in this section verify the results on the asymptotics in the previous section.
Interestingly, as far as the dilaton $\Phi$ is concerned the asymptotic solution obtained here agrees qualitatively
with the global numerical solution. In other words the dynamics of the
dilaton is not much affected by the presence of the non-BPS brane and
is qualitatively similar to that of the pre-big bang scenario.
Concerning $H(t)$ things are different: See Fig.~\ref{Bounce-H} for example. For large negative times $H(t)$ is well
described by the asymptotic solution described in Eq.~(\ref{asympt}).
Then near the bounce which takes place at $t\simeq 0$ 
\footnote{In fact one can always arrange the bounce to take place at $t=0$ by shifting the time variable.} 
the non-BPS brane affects $H(t)$ significantly. Then for $t\to t_0>0$ (near the
pole), where the pre big-bang singularity occurs, the Hubble constant
is well described by the solution in the pre-big bang scenario.  This
can be understood from the fact that the brane has already decayed for
$t\to t_0$.

\section{Nonsingular solutions -- Example}\label{SectionNonsingular}

As we have seen in the previous section, there is a singularity either in the future or in the past,
depending on the sign of $\dot{\Phi}$, and we expect that this
singularity may be resolved in the same way as in pre-big bang
scenarios, e.\,g.\ relying on $\alpha'$ corrections or quantum loop
corrections or alternatively using a dilaton potential (see
\cite{Gasperini:2007} for some explicit examples). As a matter of
fact, resolving this kind of asymptotic singularities is less
difficult than the big-bang singularity. In this section we will
give an example of resolution of this asymptotic singularity.

Near the singularity the curvature and the dilaton blow up, and this
suggests that a potential term of the form \(R\,\ee^{\Phi}\) in the
Lagrangian may smooth out the singularity. Here the coupling between
the Ricci scalar and the dilaton is introduced because the singularity
appears both in the curvature and dilaton. Such a term is quite likely
to appear as \(\alpha'\) correction in the open string sector, since
it has the form of a tree level correction in the open string coupling
constant and \(R\) is the natural invariant built from background
metric derivatives\footnote{A similar potential has been motivated in
  the context of string gas cosmology in \cite{Brandenberger:2007xu}
  as a Casimir-type potential.}.

Thus as an example, in which such an additional term may resolve the
singularities, we study the dynamics of the system where the action
(\ref{eq:original action}) is supplemented by a potential
\begin{equation}
\frac{1}{2\kappa_{10}^2}\int\de^{10}\,x\sqrt{-g}\ee^{-2 \Phi} R V(\Phi).
\label{suppl-potential}
\end{equation}
The equations of motion then take the form 
\begin{eqnarray}
72H^2\left(1+V(\Phi)\right)+4 \dot{\Phi}^2 - 
36H \dot{\Phi}  \, \left( 1+V(\Phi)-\frac{V'(\Phi)}{2} \right) - 
2 \kappa_{10}^2 \ee^{2\Phi} \epsilon &=& 0 \\
2\ddot{\Phi}-2\dot{\Phi}^2 + 18 H \dot{\Phi}-
\left( 9\dot{H}+45H^2 \right) \left( 1+V(\Phi)-\frac{V'(\Phi)}{2} \right) - 
\kappa_{10}^2 \ee^{2\Phi} p/2 &=& 0 \\
(\ddot{\Phi}+8H\dot{\Phi}) \left(V'(\Phi)-2 V(\Phi)-2 \right) + 
\dot{\Phi}^2 \left( V''(\Phi)-4 V'(\Phi)+4 V(\Phi)+2 \right) + \nonumber\\
\left(1+V(\Phi)\right)(36 H^2 +8\dot{H})+\kappa_{10}^2 \ee^{2\Phi} p &=& 0 \\
\dot{\epsilon}+9 H (\epsilon + p)-\dot{\Phi} p &=& 0.  
\end{eqnarray}

We require that the additional term should not spoil the bounce, namely 
this term is important only when the curvature becomes very big. For 
concreteness we choose $V = -\ee^{\Phi +5}/40$. Using the same setup as in 
the previous section, we perform the numerical analysis. 

First, let us consider the case in which the dilaton grows; in this
case we faced a future singularity. We found that the addition of a
potential (\ref{suppl-potential}) can resolve the future curvature singularity.
This is shown in Figs.~\ref{nonsingular-future}, where the dashed
(solid) curve corresponds to the case without (with) the additional
term. Here the initial conditions are chosen such that the bounce
takes place at $t=0$ (in other words we impose the initial conditions
at the bounce and extrapolate in both directions in time). In the case
without the additional term, there is a singularity, while in the
other case the universe evolves to the standard cosmological regime
where the Hubble parameter decreases. As can be seen from the plot,
the dynamics are almost the same in both cases before the Hubble
parameter gets significantly big, so that the bounce is not spoiled.
Once the Hubble parameter is large enough, the additional term smooths
out the singularity.%
\FIGURE{\subfigure[Hubble parameter \(H\)\label{nonsingular-H-future}]{\epsfig{file=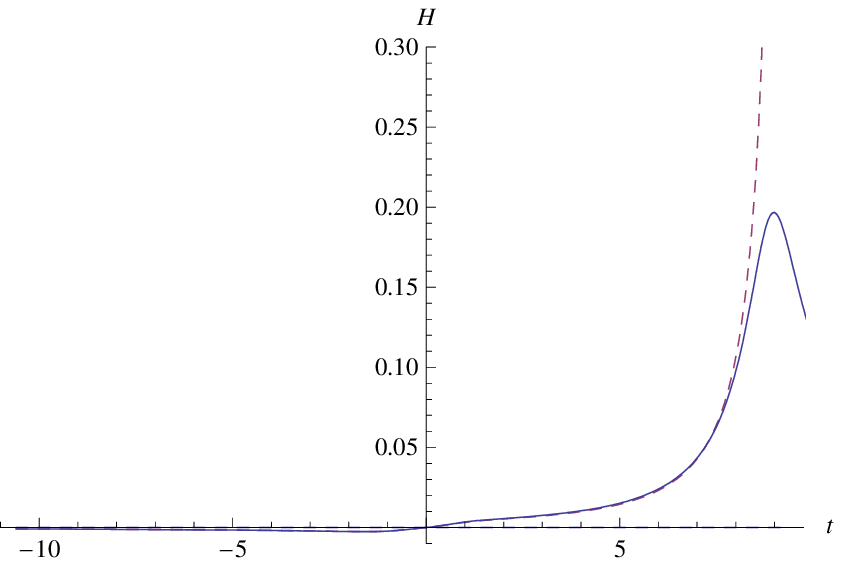,width=.475\textwidth}}
\subfigure[dilaton \(\Phi\)\label{nonsingular-Phi-future}]{\epsfig{file=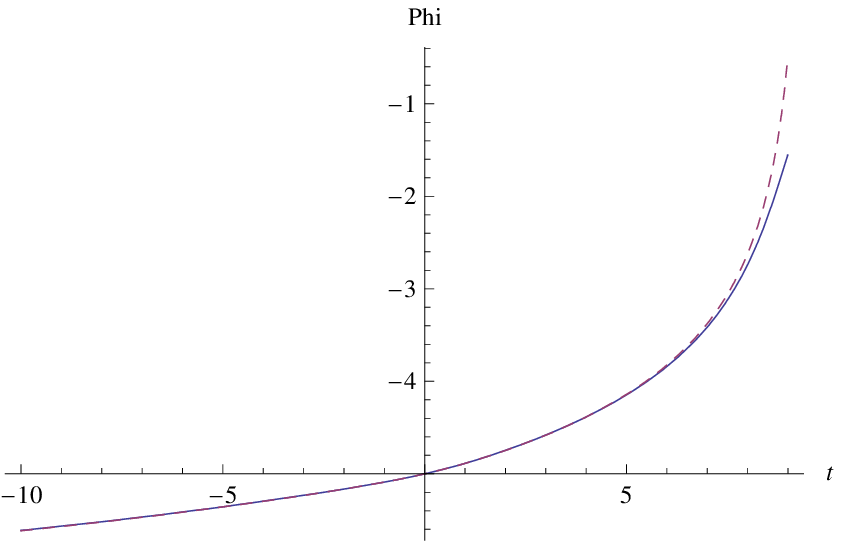,width=.475\textwidth}}
\caption{Numerical solution with growing dilaton (solid
  lines: with additional term, dashed lines: without additional term).\label{nonsingular-future}}}

Now we turn to the case in which the dilaton decreases. In the
previous section we have seen that in this case there is a past
singularity. With the help of the additional term mentioned above, we
found that this singularity can be resolved as well. This is shown in
Figs.~\ref{nonsingular-past}. The mechanism of resolving the
singularity is analogous to the case of growing dilaton that we have
seen above. What is interesting is that this solution corresponds to
the time reflected version of the case of growing dilaton since
$H\mapsto-H$ and $\dot{\Phi}\mapsto-\dot{\Phi}$ under the time reflection,
$t\mapsto-t$.
\FIGURE{\subfigure[Hubble parameter \(H\)\label{nonsingular-H-past}]{\epsfig{file=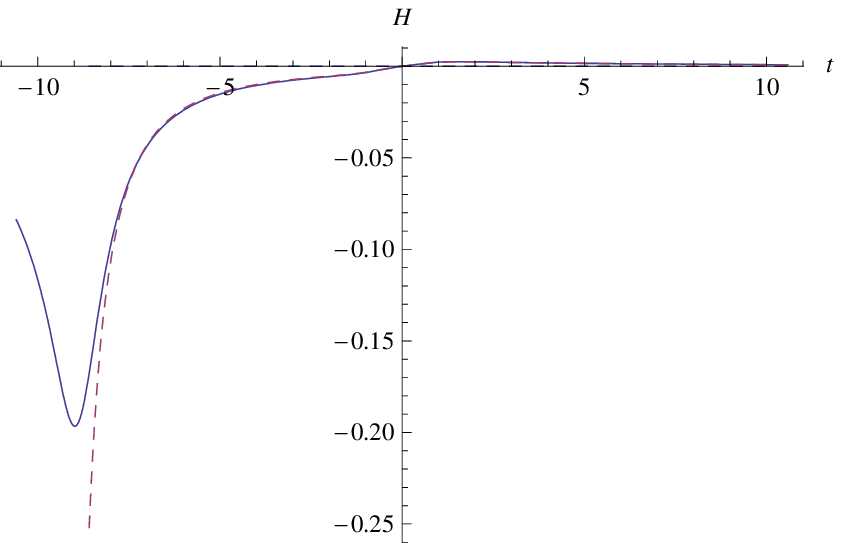,width=.475\textwidth}}
\subfigure[dilaton \(\Phi\)\label{nonsingular-Phi-past}]{\epsfig{file=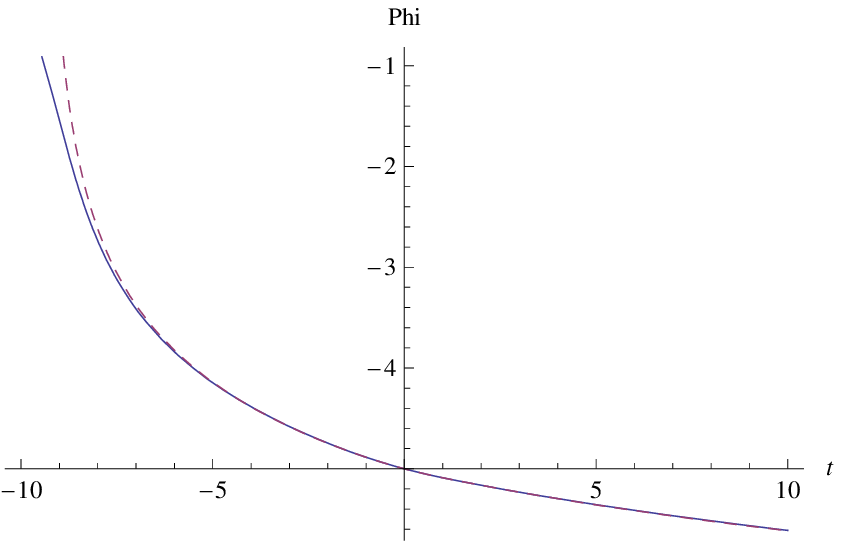,width=.475\textwidth}}
\caption{Numerical solution with decreasing dilaton (solid
  lines: with additional term, dashed lines: without additional term).\label{nonsingular-past}}}

In both cases, the dilaton dynamics still has a singularity either in the future or in the past. 
This is in contrast to the case where the tachyon sector is absent (i.\,e.\ $\epsilon=0$ and $p=0$).
(see Figs.~\ref{pre}, where the pre-big bang singularity is resolved not only in $H$, but also in $\Phi$).
Since the singular behavior of our solutions is the same as in the pre-big bang scenario, 
we expect that in principle the dilaton singularity can be resolved as in Figs.~\ref{pre}, 
but this may require fine-tuning.
\FIGURE{\subfigure[Hubble parameter \(H\)\label{H-pre}]{\epsfig{file=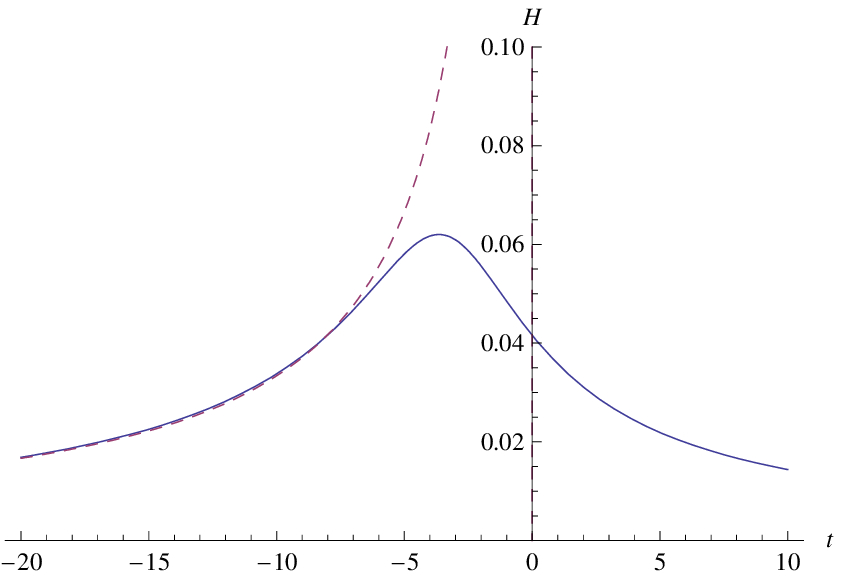,width=.475\textwidth}}
\subfigure[dilaton \(\Phi\)\label{Phi-pre}]{\epsfig{file=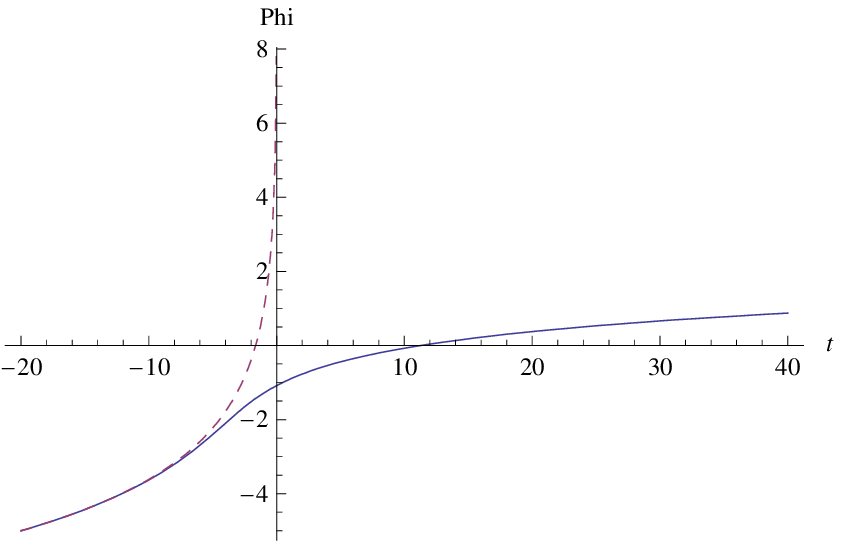,width=.475\textwidth}}
\caption{Numerical solution with tachyon sector absent (solid lines: with the potential given in (\ref{suppl-potential}), dashed lines: pre-big bang solution).\label{pre}}}

To sum up, we have shown an explicit example in which an additional
term that might arise from higher order corrections can resolve the
curvature singularity without affecting the bounce dynamics, 
though the singularity in dilaton has not been resolved.

\section{Conclusion}\label{SectionConclusion}

We suggested bounce scenarios, in which a non-BPS space-filling
D9-brane in type IIA superstring theory drives a bounce of the scale
factor in the string frame. We employed the lowest order effective
action for the metric and dilaton in the string frame as well as an
effective action for the open tachyonic mode of the non-BPS D-brane.
The positivity of the pressure of the tachyon field is responsible
for the bounce, which is why the DBI action, for instance, can not
drive the bounce.  The curvature as well as the time derivative of the
dilaton remain small during the bounce. In other words, the gravitational 
sector is entirely classical.  

Asymptotically our bounce
solutions look like pre-big bang or post-big bang solutions, with
singular behavior of the curvature and the dilaton. The asymptotic string frame
curvature singularity can be resolved by adding a phenomenological
potential, $\propto R \ee^{-\Phi}$, which may or may not result from
$\alpha'$ corrections in the open string sector. It would be desirable
to determine the sign and the precise numerical value of the
proportionality coefficient. With our choice of the sign the
gravitational coupling changes sign in the string frame. This results
in a bounce in the Einstein frame at some time after the bounce has
taken place in the string frame without violating the null energy
condition. An interesting observation is that while our
phenomenological potential stabilizes the dilaton within the
perturbative regime it fails to do so once the tachyonic sector is
included. An obvious question is then whether a modified potential
exists which stabilizes the dilaton in our model, and if so, whether
it can be derived from string theory. We should also mention that
throughout this paper we assumed the isotropy in 9-dimensional space
(modulo compactification, the details of which did not play a role
here) during the era of the bounce. For phenomenological reasons it
may be preferable to consider scenarios with a different dynamics for
the scale factor of the internal dimensions. In particular, orbifold
compactifications are interesting since they are accompanied by a
reduction of supersymmetry. A preliminary analysis shows that for a
$T^6/Z_2$ orbifold the asymptotic solutions in the far past are not
modified qualitatively. The numerical evolution of the global solution
requires more work, however. We hope to report on this issue in a
future publication.

\acknowledgments We thank S.~Hofmann for helpful comments on the
manuscript. N.~M.\ and I.~S.\ are supported in parts by the Transregio
TRR 33 `The Dark Universe' and the Excellence Cluster `Origin and
Structure of the Universe' of the DFG as well as the DFG grant, MA
2322/3-1. J.U~K. is supported by the German Academic Exchange Service
(DAAD).


\providecommand{\href}[2]{#2}\begingroup\raggedright\endgroup
\end{document}